\documentclass[12pt,preprint]{aastex}

\shorttitle{XMM observation of the SNR G337.2+0.1}
\shortauthors{Combi et al.}

\def\xmm{{\it XMM-Newton\,}}

\begin{document}


\title{XMM detection of the supernova remnant G337.2+0.1}


\author{Jorge A. Combi\altaffilmark{1}}
\affil{Departamento de F\'{\i}sica (EPS), Universidad de Ja\'en, Campus Las Lagunillas s/n, Ed-A3, 23071 Ja\'en, Spain}
\email{jcombi@ujaen.es}

\author{Juan F. Albacete Colombo\altaffilmark{2}}
\affil{Osservatorio Astronomico di Palermo, Piazza del Parlamento 1, Palermo (90141), Italy}

\author{Gustavo E. Romero\altaffilmark{2,4} and Paula Benaglia\altaffilmark{3,4}} 
\affil{Facultad de Ciencias Astron\'omicas y Geof\'{\i}sicas UNLP, Paseo del Bosque, B1900FWA La Plata, Argentina\\
Instituto Argentino de Radioastronom\'{\i}a, C.C.5, (1894) Villa Elisa, Buenos Aires, Argentina}




\begin{abstract} We report the first XMM detection of the SNR candidate
G337.2+0.1 (=AX~J1635.9$-$4719). The object shows centrally filled and diffuse
X-ray emission. The emission peaks in the hard 3.0$-$10.0 keV band. A spatially
resolved spectral study confirms that the column density of the central part of the
SNR is about N$_{H}\sim$5.9($\pm$1.5$)\times$10$^{22}$ cm$^{-2}$ and its
X-ray spectrum is well represented by a  single power-law with a photon index
$\Gamma$=0.96$\pm$0.56. The non-detection of line emission in the central
spectrum is consistent with synchrotron radiation from a population of
relativistic  electrons. Detailed spectral analysis indicates that the outer
region is highly absorbed and quite softer than the inner region, with 
N$_{H}\sim$16.2($\pm$5.2)$\times$10$^{22}$ cm$^{-2}$ and kT=4.4($\pm$2.8) keV. 
Such characteristics are already observed  in other X-ray plerions. Based on the
morphological and spectral X-ray information, we confirm the SNR nature of
G337.2+0.1, and suggest that the central region of the source is a pulsar wind
nebula (PWN), originated by an energetic though yet undetected pulsar, that is
currently losing energy  at a rate of $\sim$ 10$^{36}$ erg s$^{-1}$.  
\end{abstract}
\keywords{X-rays: individual (AX~J1635.9$-$4719) --- (ISM:) supernova remnants --- ISM: individual (G337.2+0.1) ---
X-rays: ISM --- radiation mechanisms: non-thermal}



\section{Introduction}

Supernova remnants (SNRs) of the Crab-like or plerionic class   are objects
characterized in the radio band by a compact, filled-center morphology with  a
relatively flat spectral index \citep{slane00}. In the X-ray band they present
non-thermal  spectra, characteristic of synchrotron emission. The non-thermal
spectrum, in some cases, can  even reach the very high-energy gamma-ray
region, like in MSH 15-52, G18.0-07, and Vela X \citep[see][]{aha05,aha06a}.
It is widely believed that plerions are powered by the loss of rotational
energy from energetic  pulsars, although clear evidence of the presence of
these objects is often lacking. The pulsar wind forms a nebula inside the SNR
(the PWN), where relativistic particles can be efficiently accelerated
producing synchrotron radiation that yields the typical morphologies observed
at radio and X-rays. 

The new generation of X-ray instruments with improved sensitivity and 
spatial/spectral resolution like ASCA, XMM, CHANDRA and INTEGRAL allows to
detect and  study not only distant energetic objects with these
characteristics, but also those located in regions of high density.

Recently, \citet{combi05} have presented evidence supporting a SNR origin for
the radio source G337.2+0.1. The object was discovered by the MOST galactic
plane survey at 843-MHz towards the Norma spiral arm \citep{whit96}, and
later, detected with the ASCA telescope during a survey of part of the
galactic plane \citep{sugi01}. Its integrated flux was reported to be  $\sim$
1.2$\times$10$^{-12}$ erg cm$^{-2}$ s$^{-1}$ in the 0.7-10 keV energy band.
The photon index is poorly constrained: $\Gamma$=2.8$^{+2.6}_{-1.6}$, and the
absorption column  density of the best-fit model yielded N$_{H}$=15$^{+15}_{-9}$  
$\times$ 10$^{22}$ cm$^{-2}$ \citep{combi05}. A
thorough study of its radio (continuum and line) and X-ray properties shows
that the emission from the source is consistent with what is expected for a
young SNR located at a distance $d\sim$14 kpc \citep{combi05}.  More recently,
G337.2+0.1 has been suggested as the potential counterpart  of the high-energy
gamma-ray source HESS J1634-472 \citep{aha06b}: this possibility needs further
confirmation. Throughout this paper, we adopt 14 kpc as the distance to
G337.2+0.1 (hence, 1' corresponds to  4 pc) .

In this  {\it Letter}, we present the first \xmm\ observations of the SNR
candidate G337.2+0.1. Based on its X-ray properties we are able to confirm
that this object is a non-thermal SNR with hard, featureless power-law
spectrum and, possibly, a PWN originated by a non-detected energetic pulsar.
In Section 2, we present the X-ray observations and data reduction. X-ray
analysis and results are presented in Section 3, and in Section 4 we provide a
brief discussion and a summary.

\section{X-ray Observations and data reduction}

The SNR candidate G337.2+0.1 has been marginally observed on February 2004  by
the \xmm X-ray satellite in two separated observations  (Obs-Id. 0204500201
and 0204500301). Both observations were centered towards the source
IGR\,J16358-4726 ($\alpha_{\rm J2000.0}$=\\16$^{\rm h} 35^{\rm m} 53\fs820$, 
$\delta_{\rm J2000.0}$=-47$\degr 25\arcmin 41\farcs10$), and  were acquired
with the EPIC MOS \citep{tur01} and EPIC PN \citep{stru01} cameras.
Observations were taken with a "{\it thin}" filter, and in the full frame (FF)
imaging mode.  Temporal resolution is 2.5 s and 200 ms for the MOS and PN
CCDs, 
respectively\footnote{http://xmm.vilspa.esa.es/external/xmm\_user\_support/documentation/}. 
Observations were obtained from the XMM-Newton Science Archive 
(XSA)\footnote{http://xmm.vilspa.esa.es/xsa/},  and raw EPIC data were
calibrated using the last version of the Standard Analysis System 
(SAS)\footnote{http://xmm.vilspa.esa.es/external/xmm\_sw\_cal/sas.shtml}. To
create images, spectra and light curves, we selected events with {\sc flag 0}, and 
{\sc patterns}$\leq$ 12 and 4 for MOS and PN cameras,  respectively. 

The off-angle of G337.2+0.1 respect to the center of the  observation is about
$\sim$6.67 arcmin, which implies a reduction of the  \xmm\, effective area of
about 12\%.  The net exposure times of the observations are 34.7 (Obs-Id.
0204500201) and 32.6 (Obs-Id. 0204500301) ks.  Unfortunately, the first
observation was affected by a high and variable soft proton background level
\citep{lumb02}, whereas the second one (Id. 0204500301) is  unaffected by
background fluctuations. We derive Good Time Intervals (GTI)  by the
accumulation of background light-curves in the 10-15 keV energy band, which
leads to a reduction of $\sim$87\% in the net exposure time of the Obs-Id.
0204500201. In order to avoid contamination for high background patterns
hereafter our analysis concerns only  to the observation 0204500301. The
number of detected counts in the 0.5-2.5 and 2.5-10.0 keV energy  bands are
117/121/315 and 431/403/1154 for the MOS1, MOS2 and PN cameras, respectively.
Finally, at the SNR G337.2+0.1 EPIC-PN position there is a CCD gap in the
X-ray image, leading us to ignore these data only for the X-ray image analysis
section,  but they are included for the rest of our study. 





\section{X-ray analysis of G337.2+0.1}
\subsection{Image}

The coordinates of the SNR G337.2+0.1 were defined at the position where X-ray
emission peaks ($\alpha_{\rm J2000.0}=16^{\rm h} 35^{\rm m} 54\fs95$,
$\delta_{\rm J2000.0}=-47\degr 19\arcmin 02\farcs2$), being the errors in R.A.
and DEC of $\epsilon_\alpha\,=\pm$2\fs1 and $\epsilon_\delta\,=\pm$3\farcs3,
respectively (at the 90\% of confidence). This position agrees well with the
previous estimate of the radio position  \citep{combi05}, but differs in
$\sim$ 50 arcsec from the ASCA coordinates.   Because of the poor spatial
resolution ($\sim$2.9 arcmin) the ASCA telescope alone is not conclusive in
resolving  the SNR G337.2+0.1. Fortunately, the availability of the 0.5-10.0
keV X-ray  observations from \xmm\,, the largest X-ray telescope so far
(1480 cm$^2$ @1.5 keV), allows to investigate with success  the X-ray nature
of this source. In fact, \xmm\, has $\sim$40 times more spatial resolution 
than ASCA, whereas the sensitivity limit is at least $\sim$3 times better. 

\begin{figure}[!h] 
\centering
\includegraphics[width=7.4cm,angle=0]{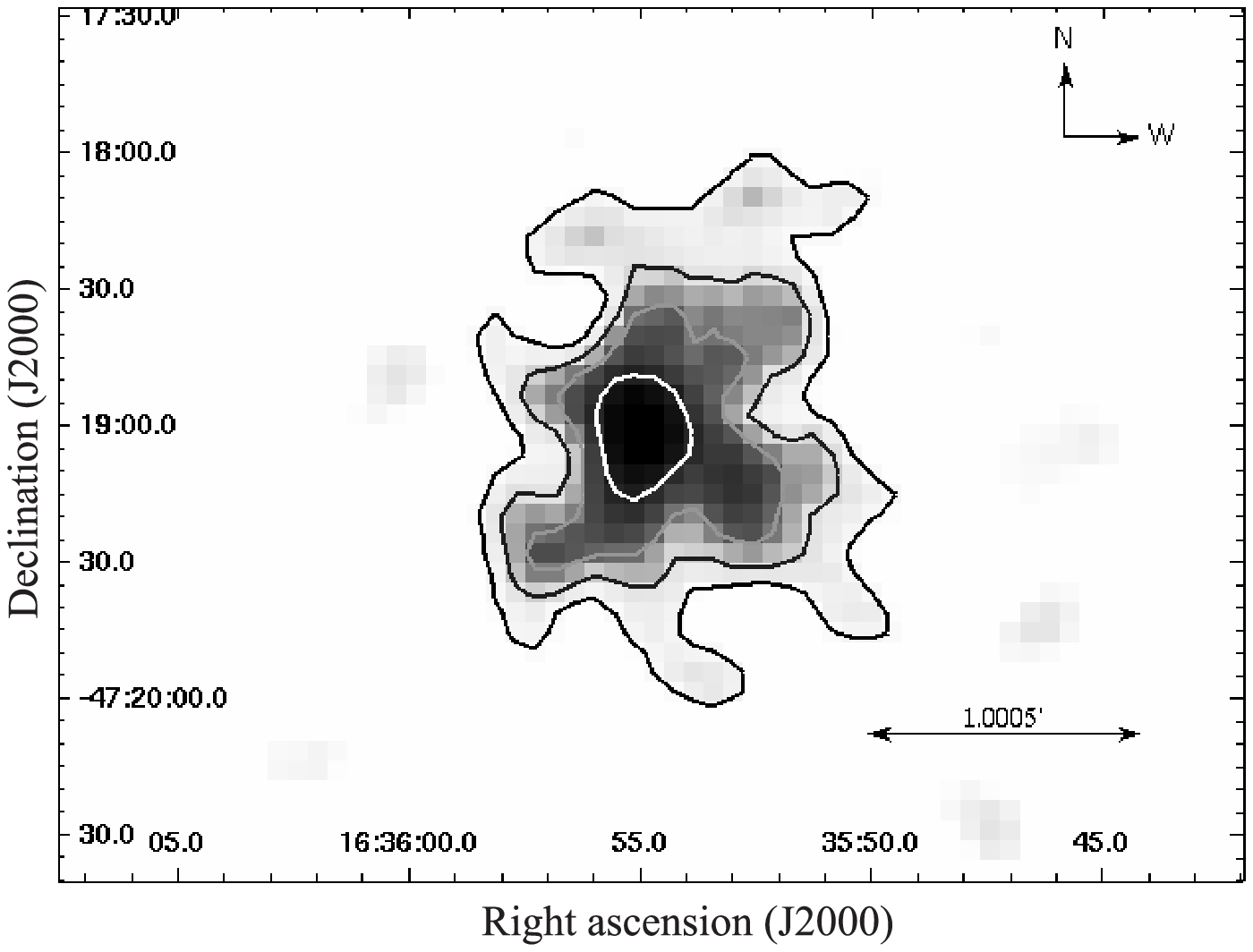}
\caption{X-ray image of the G337.2+0.1 in the 0.5-10.0 keV band. 
Typical local background level of the smoothed image is $\sim$0.94($\pm$0.17) counts\,px$^{-1}$.
Contours show the level of 1.6 ($\sim$1$\sigma$), 3.2
($\sim$2$\sigma$), 4.7 ($\sim$3$\sigma$) and 6.3 ($\sim$4$\sigma$) photons\,px$^{-1}$,
from outer to inner curves, respectively.
}
\label{imgX1}%
\end{figure}

Figure \ref{imgX1} shows the X-ray image of the SNR G337.2+0.1 in the 0.5-10.0
keV energy band. The image does not reveal a typical rim-brightened outer SNR
shell, so the overall size of the diffuse X-ray emission is uncertain.

We use the clean event files to generate MOS1 ans MOS2 images in the  energy
band [0.5-10] keV with a spatial binning of 4.35 arcsec per pixel. In order to
increase the signal-to-noise (S/N) ratio, we use the {\it emosaic}  SAS task
to merge together the two images.  The corresponding set of exposure maps for
each camera has been prepared to  account for spatial quantum efficiency and
mirror vignetting  by running the SAS task {\it eexmap}.  Exposure vignetting
corrections were performed by dividing the superposed count image by the
corresponding superposed exposure maps.  We adaptively smoothed this image to
a S/N ratio of 10 using the SAS task {\it asmooth} (see Figure\,\ref{imgX1}).
Plotted contours correspond to 1, 2, 3 and 4  sigma levels over the mean
background flux of the image ($\sim$6$\times$10$^{-5}$ ph/cm$^2$/s).

\begin{figure}[!t] 
\centering
\includegraphics[width=6.8cm,angle=0]{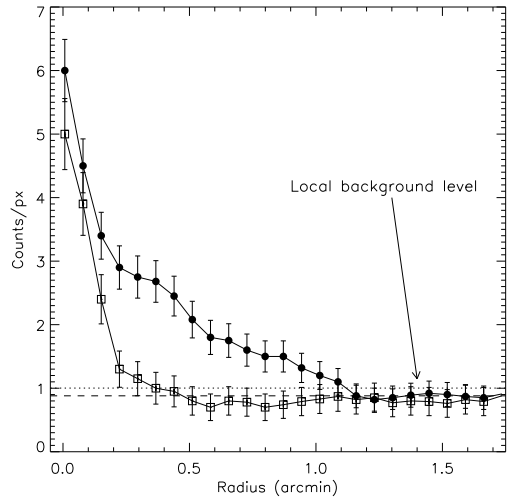}
\caption{Radial profile of the smoothed X-ray image of SNR G337.2+0.1\, ({\it filled circles}). 
Open squares refer to the radial profile of an observed point
source located at roughly the same off-axis position.}
\label{radial}%
\end{figure}

Finally, we are able to investigate the spatial extent of G337.2+0.1.
Initially, this was performed by \cite{combi05} using  ASCA data. However, the
low resolution of this image led these authors to consider a radial analysis
at large angular distance ($\sim$ 10 arcmin).   Thanks to the high resolution
\xmm\, image, we found at least 19 weak point-like X-ray  sources inside this
radius, making the ASCA studies of G337.2+0.1 to be hardly biased by source
contamination effects. According to the image presented in Figure \ref{imgX1},
G337.2+0.1 does not extend farther than 1.5 arcmin from the central peak. We
also compare G337.2+0.1 spatial extent with that produced by a point  source
placed at a similar off-axis ($\sim$ 6 arcmin) position.  In Figure
\ref{radial} we show that the SNR G337.2+0.1 has  an extension $\sim$3.5 times
larger than what is expected for a  point-like source.

\subsection{Spectral analysis}

\begin{figure*} 
\centering
\includegraphics[width=3.5cm,angle=270]{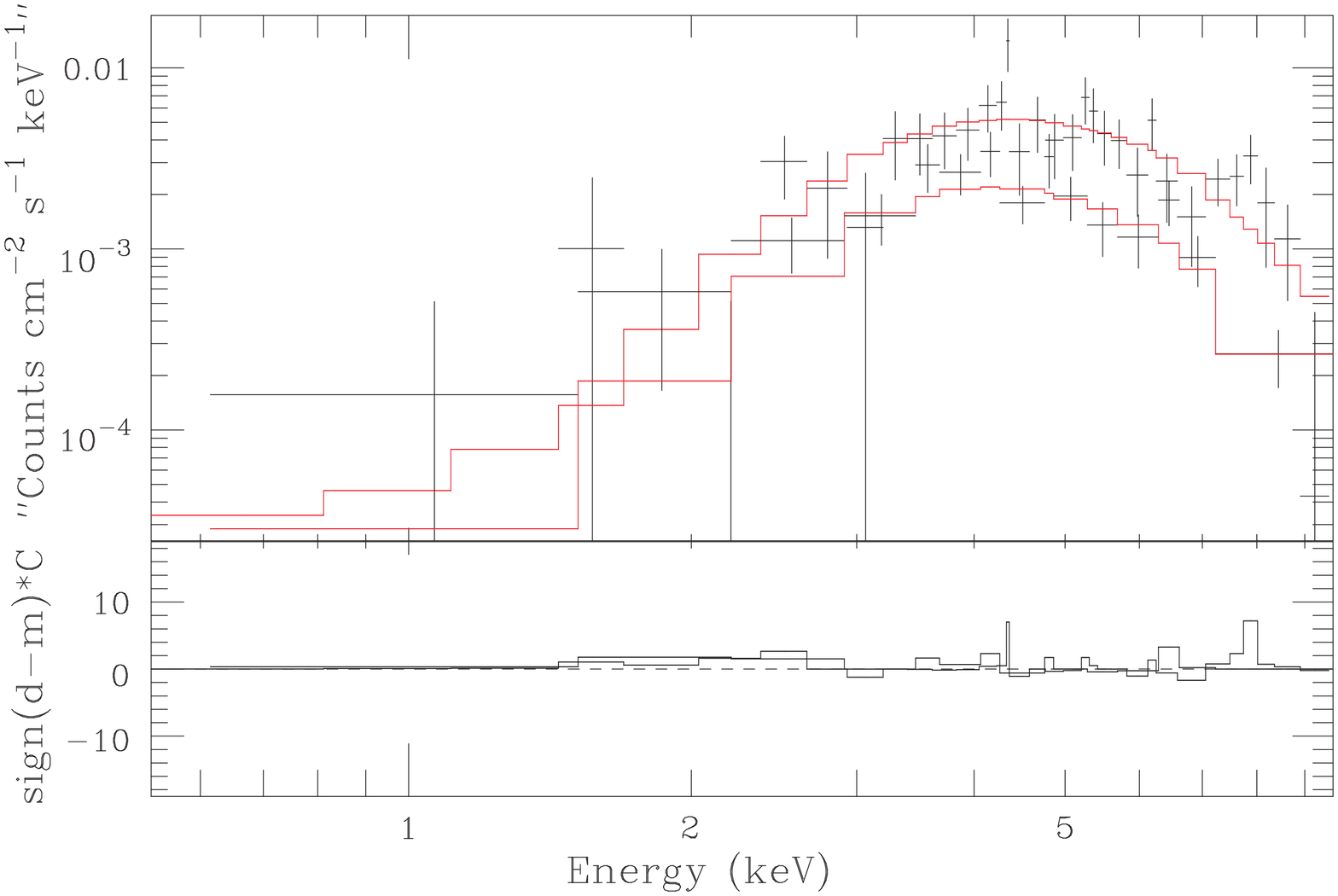}
\includegraphics[width=3.5cm,angle=270]{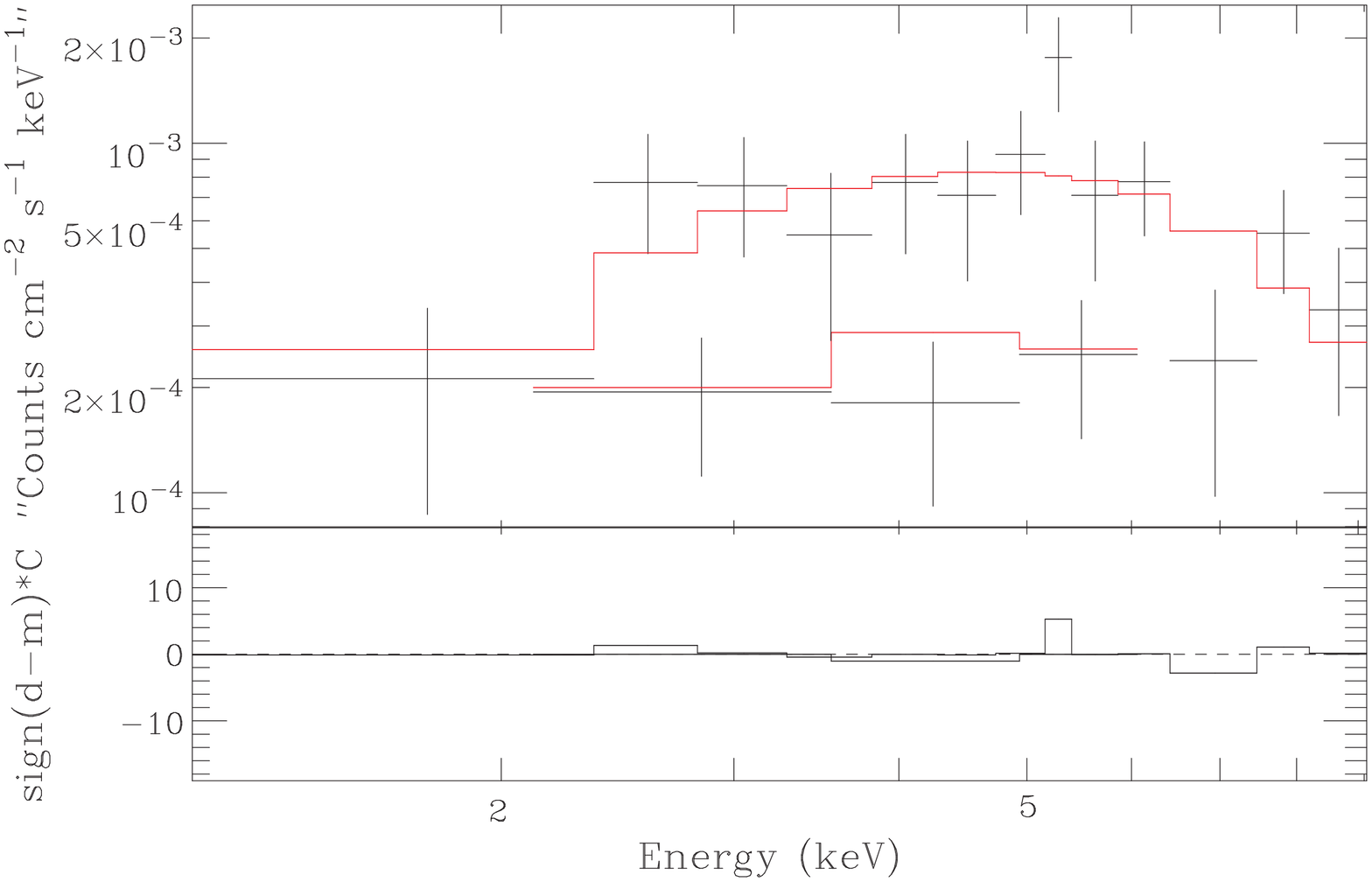}
\includegraphics[width=3.5cm,angle=270]{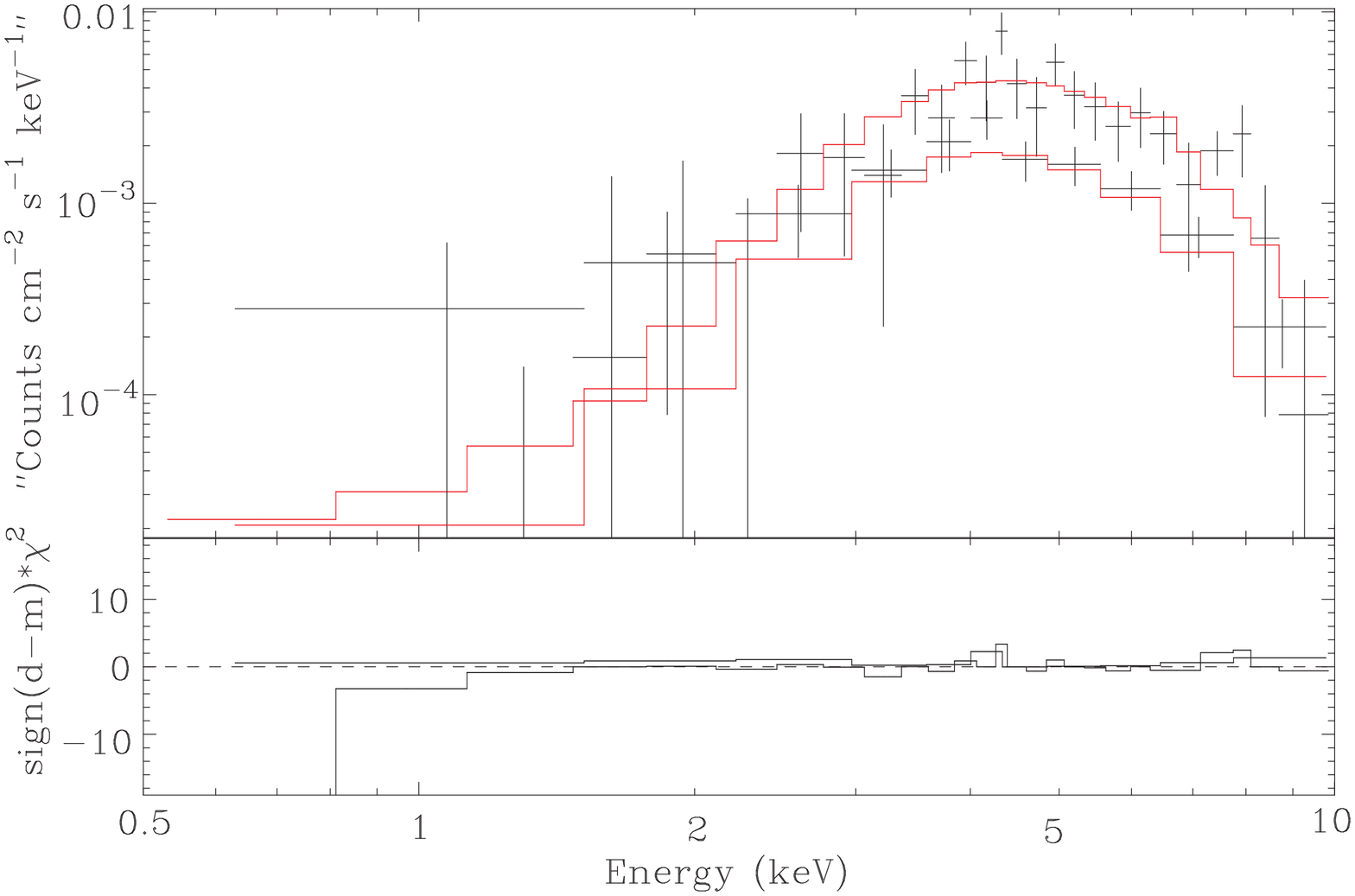}
\caption{X-ray spectra of the SNR G337.2+0.1 in the 0.5-10.0 keV
band. In all figures PN data are the upper line. Left: Spectrum extracted on the whole SNR. 
Center: Spectrum of the compact source observed in the center of SNR G337.2+0.1.
Right: Outer spectrum that excludes the central contribution.}
\label{spectra}%
\end{figure*}

For the spectral analysis we used MOS and PN data. It  was performed using the
{\sc xspec} package \citep{arnaud96}. Since the statistics of the source is not
complete enough to perform a spatial-spectral analysis, we extracted X-ray
photon events from only three different regions: $i)$ a circular region of 50
arcsec; $ii)$ a circular region of only 12 arcsec that accounts for the central
source observed in the image; $iii)$ an annulus for the extended emission of
inner-outer radii of 12-50 arcsec. The background region was taken from a
nearby blank region in the neighborhood of the source. Ancillary response files
(ARFs) and redistribution matrix files (RMFs) were calculated. All spectra are
grouped with a minimum of 16 counts per bin. The background-subtracted spectra
of the MOS and PN data (the upper line) are shown in Figure \ref{spectra}. 
We checked for possible background contamination in our spectra, inconsistency 
in the extraction of local backgrounds and differences in background corrected
spectra of SNR G337.2+0.1. We note excesses in spectra, essentially at
energies between 7.5 to 8 keV. However, such features have a low statistical 
significance ($\sim$ 1 to 1.5 sigma) and are related to fluorescence lines in
the background spectrum of the \xmm\, \citep[e.g.][]{DeLuca04}.

Our analysis of the \xmm\, EPIC spectra was essentially performed using a
single non-thermal model, described by a simple Power-Law emission model.  We
also fit the spectra by a thermal emission model \citep[{\sc
apec},][]{brick03}. Both models
were affected by an absorption ISM component \citep[{\sc wabs},][]{morri83}.
The goodness of the model fit was derived according to the $\chi^2$-test
statistics. The best fitting parameters of the models are shown in Table 1, 
and the errors quoted are 90\% confidence limits.

\begin{table*}
\label{model}
\begin{center}
\caption{X-ray spectral parameters of the SNR G337.2+0.1}
\begin{tabular}{l | ll | ll | ll}
\hline
Region:
&\multicolumn{2}{c|}{All}&\multicolumn{2}{c|}{Inner}&\multicolumn{2}{c}{Outer}
\\
\cline{2-3}  \cline{4-5} \cline{6-7}
Model:&\multicolumn{1}{|c}{Power-law}&\multicolumn{1}{c}{APEC}&
       \multicolumn{1}{|c}{Power-law}&\multicolumn{1}{c}{APEC($\dag$)}&
       \multicolumn{1}{|c}{Power-Law}&\multicolumn{1}{c}{APEC}\\
\hline
N$_{H}$ (cm$^{-2}$) in [10$^{22}$]&11.49$\pm$2.73   &    11.17$\pm$2.18  & 
                                         5.9$\pm$1.52   &     6.45$\pm$2.31  &
                                       16.21$\pm$5.62   &    16.26$\pm$5.25 \\
$\Gamma$ $|$ kT(keV)                  & 1.82$\pm$0.45   &    10.01$\pm$6.49 & 
                                        0.96$\pm$0.56   &    (64$\pm$64)   & 
                                        2.38$\pm$0.78   &     4.38$\pm$2.79  \\
Abundance$^\ddag$                     &     $---$       &       $<$0.05     &
                                            $---$       &      $<$0.05     &
                                            $---$       &      $<$0.05\\
Norm. in [$\times$10$^{-4}$]          & 4.91$\pm$1.93 &     8.28$\pm$2.83  & 
                                       0.55$\pm$0.51  &    (0.83$\pm$2.4)& 
                                        4.01$\pm$1.44   &      6.7$\pm$2.81 \\
Flux (erg\,s$^{-1}$\,cm$^{-2}$)&11.7$^{12.9}_{8.8}$ & 9.81$^{11.1}_{6.2}$ & 
                                   1.3$^{2.1}_{1.2}$  & 1.1$^{1.4}_{0.9}$ &
                                   8.2$^{9.9}_{7.2}$ & 7.4$^{9.3}_{6.8}$\\
Reduced  $\chi_\nu^2$              	&1.13         &  1.14   &
                                   	1.10 	      &  1.25 	& 
  	                                1.09 	      &  1.06   \\
\hline
\end{tabular}
\end{center}

{\bf Notes: Flux is absorption-corrected  in units of 10$^{-13}$
(erg\,s$^{-1}$\,cm$^{-2}$) calculated in the (0.5-10.0 keV) energy range. 
Normalization was calculated according to: 10$^{-14}$/4$\pi\,r^2(1-z)^2
\int\,n_e\,n_H dV$. $\dag$ APEC thermal model could not fit the spectra in  a
consistent way, yielding very ill-constrained parameters as solution.  The
non-thermal (power-law) fit seems to be the most representative emission model
for the central part of the SNR. $\ddag$ The abundance was adopted from
\citet{anders89} and left as free parameter along all our fits, but its value
is hardly affected by the low statistic of the spectra.} \end{table*}

According to the results presented in Table 1, the central part of the SNR
appears quite harder ($\Gamma \sim$ 0.96) than the outer one ($\Gamma \sim$ 2.38). We suggest that the most reasonable interpretation of observed emission from the central part of SNR is synchrotron radiation from relativistic electrons accelerated in the vicinity of the central source of the SNR.

The softening of the spectrum toward the outer regions of the nebula is a well known effect which has been seen in other X-ray plerions (e.g., G0.9+0.1, \citealp{porquet03}; 3C58, \citealp{torii00}; G21.5-0.9, \citealp{slane00}). 

To get a statistical assessment of the X-ray variability of the SNR G337.2+0.1,
we use the 32.6-ksec EPIC-PN observation to compare the time arrival
distribution of source photons by means of the Kolmogorov-Smirnov (KS) test
\citep{press92}. We use an extraction region centered in the SNR with
a radii of 50 arcsec. The total number of photons is 1470. We see no significant 
pulsed signal with a period greater than twice the read-out time of the EPIC-PN camera in the FF mode, 
which corresponds to a Nyquist limit of 400 ms.

\section{Discussion}

The X-ray morphology of G337.2+0.1 shows a centrally peaked emission,  
surrounded by a diffuse X-ray nebula.  The lightcurve of the object does not
show any significant flux variability above 0.4\,seg, implying that at first
glance, a pulsar origin for the central contribution could be ruled out.
However, a detailed spectral analysis indicates that the outer region is softer
than the inner region, a phenomenon observed previously in several X-ray
plerions with PWN (e.g., IC443, \citealp{bocchino01}).
A spectral analysis of the central component of the SNR shows that the
X-ray spectrum is well represented by a single power-law with a photon index
$\Gamma$=0.96$\pm$0.56, a value similar to that of objects powered by an
energetic pulsar \citep{gott03}. Moreover,  the non-detection of line emission
in this spectrum is consistent with synchrotron radiation from a population of
relativistic  electrons. These facts suggest a non-thermal origin for the X-ray
emission. We therefore conclude that the system G337.2+0.1/ AX~J1635.9$-$4719
is a non-thermal SNR with, possibly, a non-detected pulsar.

Possible reasons for the non-detection of a pulsar inside the SNR  are a short
rotation period (less than 400 ms) or unfavorable geometrical conditions. The
presence of a pulsar is suggested by the central X-ray peak found inside
G337.2+0.1. In what follows we explore the possibility that there exists a
hidden pulsar-powered component (plerion) within the SNR. Using the empirical
formula derived by \cite{seward88}, $\log L_{\rm X}$(erg s$^{-1}$)=1.39 $\log
\dot{E} - 16.6$, where $L_{\rm X}$ is the X-ray luminosity of the plerion in
the 0.2-4 keV band, we can make an estimate of the spin-down luminosity of the
pulsar (see, also, Becker \& Tr\"umper 1997). Using the X-ray flux of the
compact source and its nebula, $F_{\rm X}$(0.2-4 
keV)=4.9$\pm$1.7$\times$10$^{-13}$ erg s$^{-1}$ cm$^{-2}$, we get  $L_{\rm
X}$=1.1$\times$10$^{34}$ erg s$^{-1}$ (unabsorbed). This implies a spin-down
luminosity of $\dot{E} \sim$2.5$\times$10$^{36}$ erg s$^{-1}$, and a period of
$P\geq$0.08 $(t_{3} \dot{E_{ 38}})^{1/2}$(s), where $\dot{E_{38}}$ is the
spin-down luminosity in units of 10$^{38}$ erg s$^{-1}$, and $t_{3}$ is the
pulsar age in units of 10$^{3}$ years. In order to compare this result with those 
obtained with other empirical relations between the X-ray luminosity and the rate 
of the spin-down energy loss, we have used the \citet{becker97} and \citet{possenti02} 
equations. In the first case, (taking into account only the X-ray flux of the point source 
in the 0.1-2.4 keV band) the spin-down luminosity is 
$\dot{E} \sim$3$\times$10$^{35}$ erg s$^{-1}$, a factor 9 lower than the value obtained 
with the \citet{seward88} relation. In the second case (using the X-ray flux of the 
compact source and its nebula in the 2-10 keV band), we got $\dot{E} \sim$8$\times$10$^{36}$ erg s$^{-1}$, 
a value that is a factor 3 higher than the value obtained with the \citet{seward88} relation. 
On average, a value $\sim 10^{36}$ erg s$^{-1}$ seems to be reasonable. If we assume 
a pulsar period of less than 0.4 s, we obtain an upper limit for the age of the pulsar of $t \leq$ 1000 years. 

We have seen that G337.2+0.1 does not show a rim-brightened outer SNR shell.
This could be the result of the absorption of the soft thermal emission from
the forward  shock by the very high absorbing column density. Other sources
like Crab, G21.5-0.9 \citep{slane00} and 3C58 \citep{torii00} have weak or
absent X-ray rims and all are powered by young X-ray pulsars (\citealp{murray02}; \citealp{camilo06}). 

It could be interesting to compare the characteristics of G337.2+0.1 with
3C58.  The X-ray luminosities, between 0.5 and 10.0 keV, are $\sim
4.8\times10^{34}$ erg s$^{-1}$ and $\sim 2.4\times10^{34}$ erg s$^{-1}$,
respectively. The radio luminosities, at 1 GHz, are $\sim 3\times10^{32}$ erg
s$^{-1}$ and $\sim 4\times10^{32}$ erg s$^{-1}$. We see, then, that both
sources are quite similar. We notice that the estimated age of 3C58 is $\sim
800$ yr, a value based on the association of the SNR with the supernova  1181 \citep{stephenson02}. 
The nature of this SNR is discussed by \cite{camilo06}, who, however,  argue against the association 
with this historical SN \citep{bietenholz01}. The most significant difference seems to be the absence 
of any thermal component in the case of G337.2+0.1. If we compare with the Crab, on the contrary, 
we see that the Crab pulsar is injecting around two orders of magnitude more energy per time unit in the nebula than G337.2+0.1.  
The spin-down luminosity inferred for the pulsar from the new X-ray data sets
an upper limit to the energy available for high-energy cooling channels like
inverse Compton scattering and proton-proton interactions. The luminosity of
the nearby HESS source J1634-472 ($E>1$ TeV), if it is located at the same
distance inferred for G337.2+0.1, would be $L_{\gamma}\sim 7\times 10^{34}$ erg
s$^{-1}$. So, a physical association would be possible only if $\sim 7$ \% of
the spin-down luminosity is converted in high-energy $\gamma$-rays. 

Complementary studies of the PWN scenario will involve high-resolution X-ray
observations with Chandra satellite, and radio observations with ATCA, to allow
the comparison of the X-ray spectrum and morphology with those at the radio
band.  GLAST observations could reveal a GeV $\gamma$-ray source if the
proposed association with HESS J1634-472 is correct.



\acknowledgments

We thank an anonymous referee for detailed and very
constructive comments on the manuscript. J.A.C. is a researcher of the
programme {\em Ram\'on y Cajal} funded jointly by the Spanish Ministerio de
Ciencia y Tecnolog\'{\i}a and Universidad de Ja\'en.  J.A.C. acknowledges
support by DGI of the Spanish Ministerio de Educaci\'on y Ciencia under grant
AYA2004-07171-C02-02, FEDER funds and Plan Andaluz de Investigaci\'on of Junta
de Andaluc\'{\i}a as research group FQM322.  J.F.A.C acknowledges support by the
Marie Curie Fellowship Contract N$^{\circ}$ MTKD-CT-2004-002769 on the project
"The influence of Stellar High  Energy Radiation on Planetary Atmospheres". 
G.E.R. and P.B. were supported in part by grant BID 1728/OC-AR PICT 03-13291
(ANPCyT) and CONICET (PIP 5375).
\clearpage

\clearpage

\clearpage

\end{document}